\documentstyle[12pt,epsfig,fancybox]{article}
\newlength{\dinwidth}                       
\newlength{\dinmargin}                      
\def\lsim{\mathrel{\rlap{\lower4pt\hbox{\hskip1pt$\sim$}}
    \raise1pt\hbox{$<$}}}                
\def\gsim{\mathrel{\rlap{\lower4pt\hbox{\hskip1pt$\sim$}}
    \raise1pt\hbox{$>$}}}                
\newcommand{\ccbar}{c$\overline{\mbox{c}}$}

\newcommand{\km}{{\rm, }}
\newcommand{\intl}{$\int L dt $ }

\newcommand{\picb}{pb$^{-1}$}

\newcommand{\dzero}{$D^{o}$}

\newcommand{\dodob}{$D^0 \leftrightarrow \bar{D^0} $}
\newcommand{\doee}{$D^0 \rightarrow e^+ e^- $}

\newcommand{\domm}{$D^0 \rightarrow \mu^+ \mu^- $}

\newcommand{\dpmm}{$D^+ \rightarrow \pi^+ \mu^+ \mu^- $}

\newcommand{\dpee}{$D^+ \rightarrow \pi^+ e^+ e^- $}

\newcommand{\drga}{$D \rightarrow \rho \gamma $}

\newcommand{\dopp}{$D^0 \rightarrow \pi^+ \pi^- $}
\newcommand{\dokk}{$D^0 \rightarrow K^+ K^- $}
\newcommand{\dokp}{$D^0 \rightarrow K^- \pi^+ $}
\newcommand{\etal}{{\it et al.,} }
\begin{document}

\begin{flushright}
{\bf  ETHZ-IPP-PR  96-03}\\
{\bf  September 1996} \\
\end{flushright}

\vspace*{0.4cm}
\begin{center}  \begin{Large} \begin{bf}
 Rare $D$ Meson Decays at HERA\\
  \end{bf}  \end{Large}
  \vspace*{5mm}
  \begin{large}
Christoph Grab$^{\star}$\\ 
  \end{large}
\end{center}
\hspace*{.8cm} $^{\star}$ Institute for Particle Physics, ETH Z\"urich, CH-8093 Z\"urich \\
\begin{quotation}
\noindent
{\bf Abstract:}
A status report on the prospects of measuring rare decays of 
charmed mesons at HERA is given. Based on actual experience with
measuring charm at HERA, the sensitivity on limits of rare
decays is estimated.
\end{quotation}
%

\section{Why should we study rare charm decays? }
At HERA recent measurements of the charm production cross section 
in $e p$ collisions at an
energy $\sqrt{s_{ep}} \approx 300$ GeV yielded a value of about
$1 \mu$b \cite{dstar-gp}.
For an integrated luminosity of 250 \picb,
one expects therefore about $25 \cdot 10^7$ produced \ccbar \ pairs,
mainly through the boson-gluon fusion process.
This corresponds to a total of about 
 $30 \cdot 10^7$ neutral \dzero,
 $10 \cdot 10^7$ charged $D^{\pm}$,
some  $5 \cdot 10^7$ $D_S$,
and about  $5 \cdot 10^7$ charmed baryons.
A sizable fraction of this large number of $D$'s is accessible
via decays within a HERA detector, and thus 
should be used to improve substantially our knowledge on
charmed particles.

There are several physics issues of great interest. 
This report will cover however only aspects related
to the decay of charmed mesons in rare decay channels, and
in this sense provides an update of the discussion
presented in an earlier workshop on HERA physics \cite{hera-91a}.
In the following we shall discuss these aspects, and 
point out the theoretical expectations. 
Based on experiences made at HERA with charm studies,
we shall present an estimate on the sensitivity 
for the detailed case study of  the search for the
rare decay \domm.
Other challenging aspects such as the production mechanism
and detailed comparisons with QCD calculations, or the use
of charmed particles in the extraction of proton and photon
parton densities, will not be covered here.

Possibly the most competitive future source of $D$-mesons is
the proposed tau-charm factory. 
The continuing efforts 
at Fermilab (photoproduction and hadroproduction experiments),
at CERN (LEP) and at Cornell(CESR), 
which are presently providing the highest
sensitivities, are compared with the situation at HERA.
In addition, all these different approaches 
provide  useful and complementary information 
on various properties in the charm system.


\section{Decay processes of interest}

\subsection{Leading decays }

The charm quark is the only heavy quark besides the b quark and can be used
to test the heavy quark symmetry \cite{rf-isgurw} 
by measuring form factors or decay constants.
Hence, the $D$-meson containing a charmed quark is heavy as well
and disintegrates through a large number of decay channels. 
The leading decays
$c \rightarrow s +  q{\bar q}$ or 
$c \rightarrow s + {\bar l} \nu$ 
occur with branching ratios of order a few \% 
and allow studies of QCD mechanisms
in a transition range between high and very low energies.

Although experimentally very challenging, the search for
the purely leptonic decays 
$D^{\pm} \rightarrow \mu^{\pm} \nu$ and an improved 
measurement of $D_S^{\pm} \rightarrow \mu^{\pm} \nu$
should be eagerly pursued further, 
since these decays
offer  direct access to the meson decay constants $f_D$ and $f_{D_S}$,
quantities that can possibly be calculated accurately by lattice
gauge theory methods
\cite{rf-marti},\cite{rf-wittig}.

\subsection{ Singly Cabibbo suppressed decays (SCSD)}
Decays suppressed by a factor $\sin{\theta_C}$, the socalled
singly Cabibbo suppressed decays (SCSD),
are of the form
$c \rightarrow d u {\bar d}$ or 
$c \rightarrow s {\bar s} u$.
Examples of SCSD, such as 
$D \rightarrow \pi \pi$ or $ K \bar{K}$ have been observed
at a level of $10^{-3}$ branching ratio 
(1.5 and 4.3 $\cdot 10^{-3}$, respectively)
 \cite{rf-partbook}.
They provide information about the
CKM-matrix, and also are background
processes to be worried about in the search for rare decays.

\subsection{ Doubly Cabibbo suppressed decays and
$D^0 \longleftrightarrow {\bar D^0}$ mixing}

Doubly Cabibbo suppressed decays (DCSD) of the form
$c \rightarrow d {\bar s} u$ have 
not been observed up to now\cite{rf-partbook},
with the exception
of the mode  $BR(D^0 \to K^+ \pi^- )$ that has a branching
ratio of $(2.9 \pm 1.4) \cdot 10^{-4}$.
The existing upper bounds are at the level of a few $10^{-4}$,
with branching ratios expected at the level of $10^{-5}$.
These DCSD are particularly interesting from the QCD-point of view, 
and quite a few predictions have been made\cite{rf-bigi}.
DCSD also act as one of the main background processes
to the \dodob \ mixing and therefore must be well understood,
before the problem of mixing itself can be successfully attacked.

As in the neutral Kaon and B-meson system, mixing between the 
$D^0$ and the $\bar{D^0}$ is expected to occur (with $\Delta C = 2$).
The main contribution is expected due to long distance effects,  estimated
to be as large as about
$r_D \sim 5 \cdot 10^{-3}$
\cite{rf-wolf},
while the standard box diagram yields $r_D \sim 10^{-5}$
\cite{rf-chau}.
Here $r_D$ is the mixing parameter
$ r_D \simeq (1/2) \cdot ( \Delta M / \Gamma)^2 $, with contributions by the
DCSD neglected.
Recall that the DCSD poses a serious background source in case
only the
time-integrated spectra are studied. The two sources can however be
better separated,
if the decay time dependence of the events is recorded separately
(see e.g. \cite{rf-anjos}). More details on the prospect of
measuring mixing at HERA are given in \cite{yt-mixing}.

\subsection{ Flavour Changing Neutral Currents   (FCNC)}

\begin{figure}[ht]
\epsfig{file=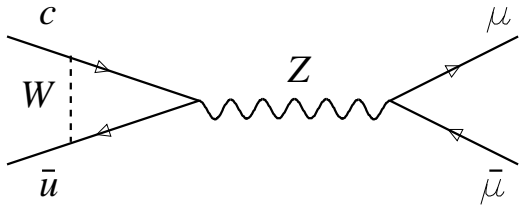,width=9cm}
\caption{\it Example of an FCNC process in the standard model  
at the loop level: \domm\ . }
\label{feyn-loop}
\end{figure}

\begin{figure}[ht]
\epsfig{file=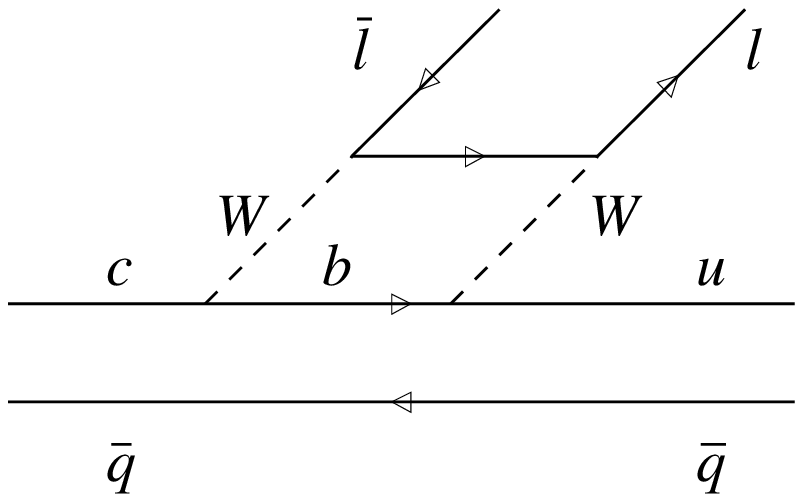,width=7.5cm}
\epsfig{file=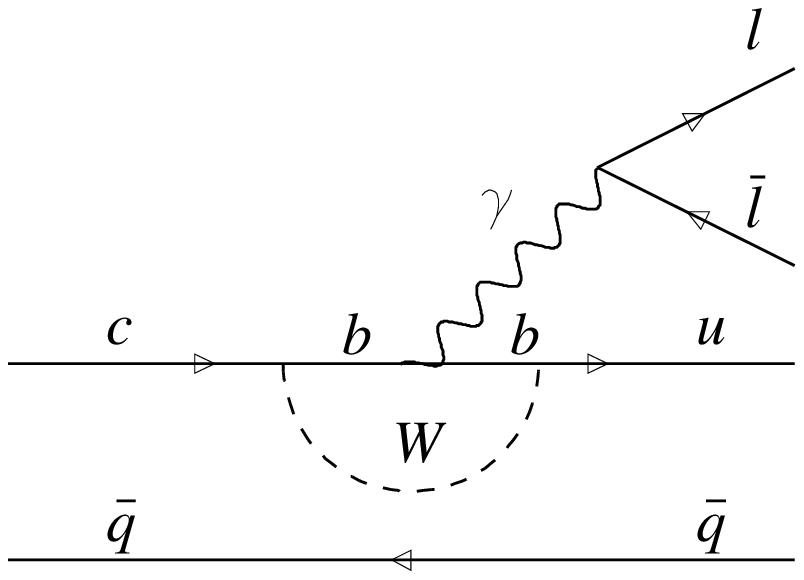,width=7.5 cm}
\caption{\it FCNC processes: short range contributions  due to
 box diagrams (a) or penguin diagrams (b).}
\label{feyn-penguin}
\end{figure}

An important feature of the standard model is that {\it flavour 
changing neutral currents (FCNC with $\Delta C =1$)} 
only occur at the one loop level in the SM
{\it i.e.} through  short distance contributions,
such as e.g. in penguin and box diagrams
as shown in figs.\ref{feyn-loop} and
\ref{feyn-penguin}.
These are transitions of the form
$s \rightarrow d + N$ or
$c \rightarrow u + N$, where $N$
is a non-hadronic neutral state such as $\gamma \ $ or $\ l{\bar l}$, and give 
rise to the decays
\drga, \domm, \dpmm \ etc. 
Although the relevant couplings are the same as those of leading decays, 
their rates are very small as they are suppressed by
the GIM mechanism \cite{gim} and the unfavourable quark masses 
within the loops.
The SM-prediction for the branching ratios are
of order $10^{-9}$ for  $D^0 \to X l^+ l^-$ and
of  $O(10^{-15})$ for  $D^0 \to l^+ l^-$, due to additional
helicity suppression.
A summary of the expected branching ratios obtained from
calculations of the loop integrals (\cite{rf-willey}, \cite{rf-bigi},
\cite{hera-91a}, \cite{long-range})
using also the QCD- short distance
corrections available \cite{rf-cella} is given in 
table \ref{tab-exp}.

However, FCNC are sensitive to new, heavy particles in the loops,  and 
above all, to new physics in general.

\begin{figure}[ht]
\epsfig{file=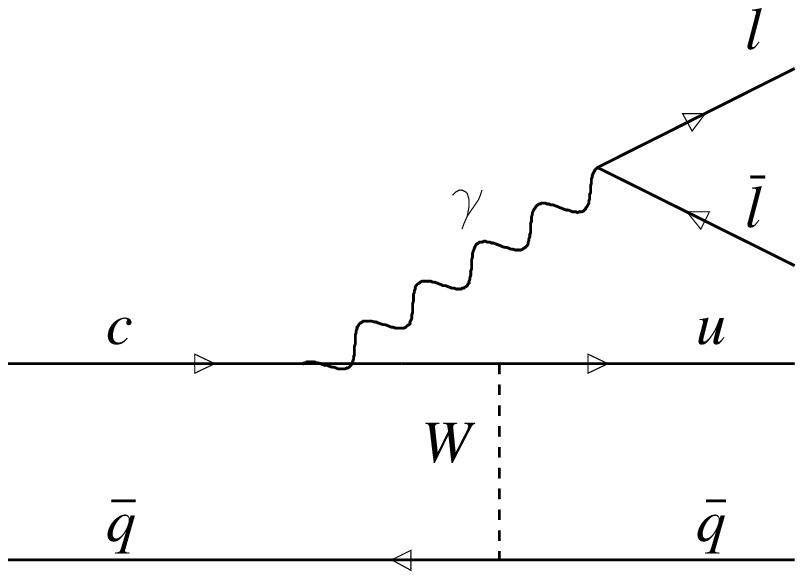,width=7.5cm}
\epsfig{file=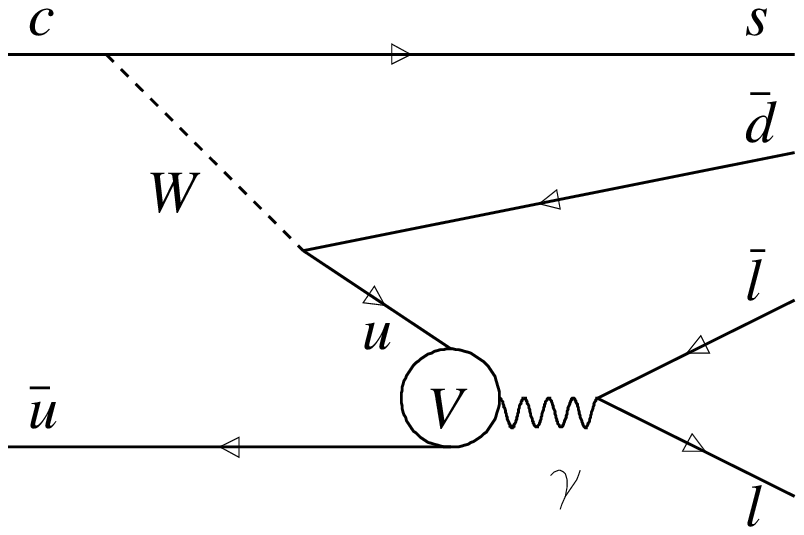,width=7.5cm}
\caption{\it FCNC processes : long range contributions due to 
$\gamma$-pole amplitude (a) and vector meson dominance (b).}
\label{feyn-gpole}
\end{figure}

In addition to these short distance loop diagrams, there are contributions
from long distance effects, which might be even larger by several
orders of magnitude\cite{long-range}. 
To mention are photon pole amplitudes 
($\gamma$-pole)
and vector meson dominance (VMD) induced processes.
The $\gamma$-pole model (see fig.\ref{feyn-gpole})
in essence is a W-exchange decay with a 
virtual photon radiating from one of the quark lines. The behaviour
of the amplitude depends on the spin state of the final state
particle (vector V or pseudoscalar P).
The dilepton mass distribution  for
$D \to V l^+ l^-$ modes peaks at zero (small $Q^2$) since the photon
prefers to be nearly real. On the other hand, the pole amplitude
for $D \to P l^+ l^-$ decays vanishes for small dilepton masses
because  $D \to P \gamma$ is forbidden by angular momentum
conservation.
The VMD model (see fig.\ref{feyn-gpole}b) proceeds through the
decay $D \to X V^0 \to X l^+ l^-$. 
The intermediate vector meson  $V^0$  ($\rho, \omega, \phi$)
mixes with a virtual photon which then couples to the lepton pair.
The dilepton mass spectrum therefore will exhibit poles at the
corresponding vector meson masses due to real $V^0$ mesons decaying.

Observation of FCNC processes at rates that exceed the
long distance contributions hence opens a window
into physics beyond the standard model.
Possible scenarios include leptoquarks or heavy neutral leptons
with sizable couplings to $e$ and $\mu$.

A measurement of such long distance contributions in the
charm sector is inherently
of interest, as it can be used to estimate similar effects
in the bottom sector \cite{long-d},
e.g. for the decay  $b \to s \gamma$, 
which was seen at the level of $2.3 \cdot 10^{-4}$.
A separation of short and long range contributions would allow
e.g. a determination of  $\mid V_{td}/V_{ts} \mid$
from the ratio
$BR(B \to \rho \gamma) / BR(B \to K^* \gamma)$
and bears as such a very high potential.
 
\begin{table}
\begin{center}
\begin{tabular}{|l|c|}
\hline
 Decay mode & Expected branching ratio \\
\hline
\hline
%
 $ c \to u \gamma $ &  $10^{-15} - 10^{-14}$  \\
 $ D \to \rho \gamma $ &  $10^{-7}$           \\
 $ D \to \gamma \gamma $ &  $10^{-10}$  \\
\hline
 $ c \to u l {\bar l} $ &  $5 \cdot 10^{-8}$  \\
 $ D^+ \to \pi^+ e^+ e^- $ &  $10^{-8}$  \\
 $ D^0 \to \mu^+ \mu^- $ &  $10^{-19}$  \\
\hline
\hline
\end{tabular}
\caption[Expectations for loop processes.]
{Expectations for branching ratios of loop processes
based on SM calculations, hereby assuming the
BR of both   $D \to \rho \rho $ and
  $D \to \rho \pi$ to be below $10^{-3}$.}
\label{tab-exp}
\end{center}
\end{table}

\subsection{ Forbidden decays }

\begin{figure}[ht]
\epsfig{file=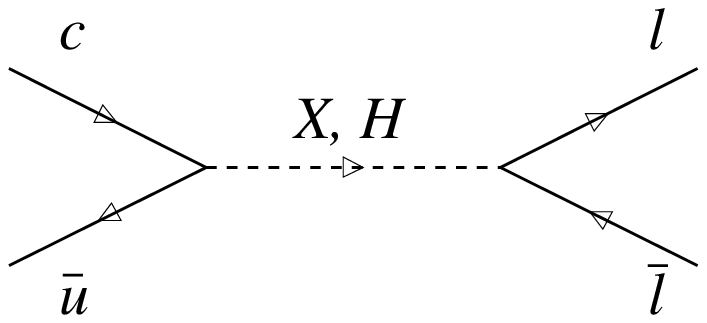,width=9.2cm}
\epsfig{file=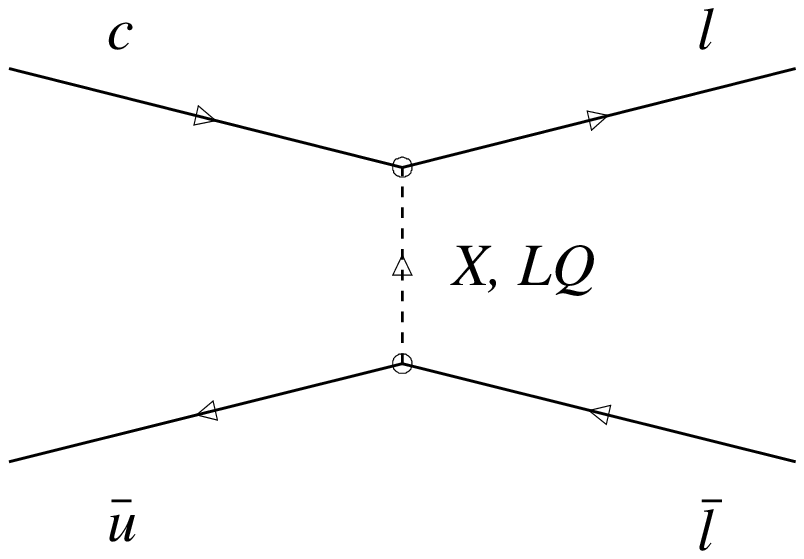,width=7.6cm}
\caption{\it FCNC processes or LFNV decays, mediated by
the exchange of a scalar particle X 
or a particle H mediating ``horizontal interactions'', or a leptoquark LQ.}
\label{feyn-x-s}
\end{figure}

Decays which are not allowed
to all orders in the standard model, the {\it forbidden} decays,
are exciting signals of new physics. 
Without claim of completeness, we shall list
here some of the more important ones:

\begin{itemize}
 \item  Lepton number (L) or lepton family (LF) number violation (LFNV) 
 in decays such as $D^0 \to \mu e$,  $D^0 \to \tau e$. 
It should be strongly emphasized that decays of $D$-mesons test 
couplings complementary to those effective in K- or B-meson decays.
Furthermore, the charmed quark is the only possible charge 2/3 
quark which allows
detailed investigations of unusual couplings.
 These are often predicted to occur in models with 
i) technicolour \cite{rf-masiero};
ii) compositeness \cite{rf-lyons};
iii) leptoquarks \cite{rf-buchmu} \cite{rf-campb}; 
(see e.g. fig.\ref{feyn-x-s}a and b); this can include
among others non-SU(5) symmetric   flavour-dependent 
couplings  (u to $l^{\pm}$, and d to $\nu$), which
would forbid decays of the sort $K_L \to \mu \mu, \ \mu e $, while
still allowing for charm decays;
iv) massive neutrinos (at the loop level) in an extended standard model;
v) superstring inspired phenomenological models 
  e.g. MSSM models with a Higgs doublet;
vi) scalar exchange particles that would manifest
  themselves e.g. in decays of the form $D^0 \to  \nu {\bar \nu}$.
%
\item Further models feature {\it horizontal} interactions,
mediated by particles connecting u and c or d and s quarks
(see e.g. fig.\ref{feyn-x-s}a).
They appear with similar
signatures as the doubly Cabibbo suppressed decays.
\item    Baryon number violating decays, such as  
   $D^0 \to p e^-$ or $D^+ \to n e^-$. They
are presumably very much suppressed,
although they are not directly related to proton decay.
\item The decay
 $ D \rightarrow \pi \gamma $ is absolutely forbidden by gauge invariance
  and is listed here only for completeness.
\end{itemize}

\vspace{-1.cm}

The clean leptonic decays make it possible to search for leptoquarks. 
If they do not couple also to quark-(anti)quark pairs, they cannot cause
proton decay but yield decays such as
$K \rightarrow \bar{l_1} l_2 $ or
$D \rightarrow \bar{l_1} l_2 $. 
In the case of scalar leptoquarks 
there is no helicity suppression and consequently the
experimental sensitivity to such decays is enhanced. 
Let us emphasize here again, that
decays of $D$-mesons are complementary to those of Kaons, since they probe
different leptoquark types. 
To estimate the sensitivity we write the effective four-fermion coupling as
$(g^2_{eff}/M^2_{LQ})$, and  obtain
\begin{equation}
 \frac{ (M_{LQ}\ /\ 1.8\ TeV)}{g_{eff}}
 \geq
  \sqrt[4] {\frac{10^{-5}}{BR(D^0 \rightarrow \mu^+\mu^-) }}.
\label{mlq}
\end{equation}
Here $g_{eff}$ is an effective coupling and includes possible mixing effects.
 Similarly, the decays $D^+ \rightarrow e^+ \nu$, \dpee \  can be used to set bounds 
on $M_{LQ}$. With the expected sensitivity, one can probe heavy exchange particles 
with masses in the  $1 \ (TeV / g_{eff})$ range.

Any theory attempting to explain the hadron-lepton symmetry or the
``generation'' aspects of the standard model will give rise to new phenomena
connected to the issues mentioned here. Background problems make it quite
difficult to search for signs
of them at high energies; therefore precision experiments
at low energies (like the highly successful $\mu$-, $\pi$- or K-decay experiments)
are very suitable to probe for any non-standard phenomena.


\section{Sensitivity estimate for HERA}

 In this section we present 
an estimate on the sensitivity to detect the
decay mode \domm.
As was pointed out earlier, this is among the
 cleanest manifestation of FCNC or LFNV processes
\cite{rf-willey}.
We base the numbers on our experience 
gained in the analysis of the 1994 data, published in \cite{dstar-gp}.
There the $D$-meson decay is measured in the decay mode
$   D^{*+} \rightarrow D^{0} \pi^{+}_s\  ; \ 
D^0 \to  K^{-}\pi^{+}$, exploiting 
the well established  $D^{*+}(2010)$ tagging 
technique\cite{rf-feldma}.
In analogy, we assume 
for the decay chain $   D^{*+} \rightarrow D^{0} \pi^{+}_s ;
 D^{0} \rightarrow \mu^+ \mu^- $, 
a similar resolution of $\sigma \approx 1.1$ MeV in
the mass difference 
$  \Delta M = M(\mu^+ \mu^- \pi^+_s) - M(\mu^+ \mu^- ) $
as in \cite{dstar-gp}.
\noindent
In order to calculate a sensitivity for the \domm 
decay branching fraction we make the following
assumptions:

\noindent
i) luminosity $L = 250\ pb^{-1} $;
ii) cross section 
 $\sigma (e p \to c {\bar c} X) \mid_{\sqrt s_{ep} \approx 300; Q^2< 0.01}
  = 940\  nb $;
iii) reconstruction efficiency $\epsilon_{reconstruction} = 0.5 $;
iv) trigger efficiency 
  $\epsilon_{trigger} = 0.6 $; this is based
 on electron-tagged events, and hence applies to
 photoproduction processes  only.
v) The geometrical acceptance $A$ has been properly calculated
  by means of Monte Carlo simulation for both 
  decay modes \dokp\  and \domm\ for a rapidity interval of
  $\mid \eta \mid < 1.5 $. For the parton density functions 
  the GRV parametrizations were employed, and the
  charm quark mass was assumed to be $m_c = 1.5$. We obtained \\
   $A = 6 \%$ for $p_T(D^*) > 2.5$ (for $K^{-}\pi^{+}$ ) \\
   $A = 18 \%$ for $p_T(D^*) > 1.5$  (for $K^{-}\pi^{+}$ ) \\
   $A = 21 \%$ for $p_T(D^*) > 1.5$  (for $\mu^+ \mu^- $) 

\noindent
A direct comparison with the measured decays $N_{K \pi}$
into $ K^{-}\pi^{+}$ \cite{dstar-gp} then yields the expected
number of events $N_{\mu \mu}$ and determines the branching ratio to
\vspace*{-0.5cm}

$$ BR(D^0 \to \mu^+ \mu^-) = BR(D^0 \to  K^{-}\pi^{+}) \cdot
  \frac{N_{\mu \mu}}{L_{\mu \mu}} \cdot \frac{L_{K \pi}}{N_{K \pi}}
  \cdot \frac{A(p_T>2.5)}{A(p_T>1.5)}  $$

Taking the numbers from \cite{dstar-gp} 
$N_{K \pi} = 119$ corresponding to an integrated
luminosity of  $L_{K \pi} = 2.77 \ pb^{-1}$,
one obtains  
\vspace*{0.5cm}

\fbox{ $BR(D^0 \to \mu^+ \mu^-) = 1.1 \cdot 10^{-6} \cdot N_{\mu \mu}$ }

\noindent
In the case of {\it NO} events observed, an upper limit
on the branching ratio calculated by means of Poisson statistics
$(N_{\mu \mu} = 2.3)$, yields a value of 
    $BR(D^0 \to \mu^+ \mu^-) < 2.5 \cdot 10^{-6}$  at 90\% c.l.

In the case of an observation of
 a handful events e.g. of $O(N_{\mu \mu} \approx 10$), one obtains
    $BR(D^0 \to \mu^+ \mu^-)  \approx 10^{-5}$.
 This can be turned into an estimate for the mass of a potential
 leptoquark mediating this decay according to eqn.\ref{mlq},
and yields a value of 
$M_{LQ}/g_{eff} \approx 1.8$ TeV.

\begin{table}[tb]
\begin{center}
\begin{tabular}{|l|c|c|c|}
\hline
 Mode & BR (90\% C.L.) & Interest & Reference \\
\hline
\hline
%
 $ r= {({\bar D^0} \to \mu^+X)} \over {({\bar D^0} \to \mu^-X)} \ $ &
$1.2*10^{-2}$ &$\ $  $\Delta C = 2$, Mix $\ $ & BCDMS 85    \\
 $ \ $ &  $5.6*10^{-3}$ &   $\Delta C = 2$, Mix  & E615 86    \\
\hline
 ${(D^0 \to {\bar D^0} \to K^+\pi^-)} \over
  {(D^0 \to K^+ \pi^- + K^- \pi^+ )}$ &
 $4*10^{-2}$ &$\ $  $\Delta C = 2$, Mix $\ $ & HRS 86   \\
 $\ $ & $ = 0.01^*$ &  $\Delta C = 2$, Mix  & MarkIII 86 \\
 $\ $ &  $1.4*10^{-2}$ &  $\Delta C = 2$, Mix  & ARGUS 87 \\
 $ \ $ & $3.7*10^{-3}$ &$\ $  $\Delta C = 2$, Mix $\ $ & E691 88   \\
\hline
 $D^0 \to \mu^+ \mu^-$ & $7.0*10^{-5}$ & FCNC  & ARGUS 88 \\
 $D^0 \to \mu^+ \mu^-$ & $3.4*10^{-5}$ & FCNC & CLEO 96 \\
 $D^0 \to \mu^+ \mu^-$ & $1.1*10^{-5}$ & FCNC & E615 86 \\
\hline
 $D^0 \to e^+ e^-$ & $1.3*10^{-4}$ & FCNC   & MarkIII 88  \\
 $D^0 \to e^+ e^-$ & $1.3*10^{-5}$ & FCNC   & CLEO 96  \\
\hline
 $D^0 \to \mu^{\pm} e^{\mp}$ & $1.2*10^{-4}$ & FCNC, LF & MarkIII 87 \\
 $D^0 \to \mu^{\pm} e^{\mp}$ & $1.0*10^{-4}$ & FCNC, LF & ARGUS 88 \\
 $D^0 \to \mu^{\pm} e^{\mp}$ & $(1.9*10^{-5})$ & FCNC, LF &  CLEO 96  \\
\hline
 $D^0 \to {\bar K}^0 e^+ e^-$ & $ 1.7*10^{-3} $ & \  & MarkIII 88  \\
 $D^0 \to {\bar K}^0 e^+ e^-/\mu^+ \mu^-/\mu^{\pm} e^{\mp}$ 
             & $ 1.1/6.7/1.*10^{-4} $ & FCNC, LF & CLEO 96  \\
 $D^0 \to {\bar K}^{*0} e^+ e^-/\mu^+ \mu^-/\mu^{\pm} e^{\mp}$ 
             & $ 1.4/11.8/1.*10^{-4} $ & FCNC, LF & CLEO 96  \\
 $D^0 \to \pi^0 e^+ e^-/\mu^+ \mu^-/\mu^{\pm} e^{\mp}$ 
             & $ 0.5/5.4/.9*10^{-4} $ & FCNC, LF & CLEO 96  \\
 $D^0 \to \eta e^+ e^-/\mu^+ \mu^-/\mu^{\pm} e^{\mp}$ 
             & $ 1.1/5.3/1.*10^{-4} $ & FCNC, LF & CLEO 96  \\
 $D^0 \to \rho^0 e^+ e^-/\mu^+ \mu^-/\mu^{\pm} e^{\mp}$ 
             & $ 1./4.9/0.5*10^{-4} $ & FCNC, LF & CLEO 96  \\
 $D^0 \to \omega e^+ e^-/\mu^+ \mu^-/\mu^{\pm} e^{\mp}$ 
             & $ 1.8/8.3/1.2*10^{-4} $ & FCNC, LF & CLEO 96  \\
 $D^0 \to \phi e^+ e^-/\mu^+ \mu^-/\mu^{\pm} e^{\mp}$ 
             & $ 5.2/4.1/0.4*10^{-4} $ & FCNC, LF & CLEO 96  \\
\hline
 $D^0 \to K^+ \pi^-\pi^+\pi^-$ & $< 0.0015$ & DC & CLEO 94   \\  
 $D^0 \to K^+ \pi^-\pi^+\pi^-$ & $< 0.0015$ & DC & E691 88   \\  
 $D^0 \to K^+ \pi^-$ & $=0.00029$ & DC & CLEO 94 \\
 $D^0 \to K^+ \pi^-$ & $< 0.0006$ & DC & E691 88 \\
\hline
\end{tabular}
\caption[Experimental limits on rare $D^0$-meson decays.]
{Experimental limits at 90\% c.l. on rare $D^0$-meson decays
(except where indicated by =).
Here L, LF, FCNC, DC and Mix denote
lepton number and lepton family number violation, flavour changing
neutral currents, doubly Cabibbo suppressed decays and mixing,
respectively.}
\label{tab-d}
\end{center}
\end{table}

\section{Background considerations}

\subsection{Background sources and rejection methods}
The most prominent sources of background originate from 
i) genuine leptons from semileptonic B- and D-decays,
      and  decay muons from $K, \pi$ decaying in the detector;
ii) misidentified hadrons,  {\it i.e.} $\pi, K$, from other
    decays, notably leading decays and SCSD; and
iii) combinatorial background from light quark processes.

The background can be considerably suppressed by applying various
combinations of the following techniques:   
\begin{itemize}
\item $D^*$-tagging technique \cite{rf-feldma}: \\
 A tight window on the mass difference $\Delta M$ is the most
  powerful criterium.
\item Tight kinematical constraints (\cite{rf-grab2},\cite{hera-91a}): \\
 Misidentification of hadronic $D^0$ 2-body decays such as
\dokp ($3.8\%$ BR), \dopp ($0.1\%$ BR) and \dokk ($0.5\%$ BR)
are  suppressed by more than an order of magnitude
by a combination of tight windows
on both  $\Delta M$ and $M^{inv}_D$.
Final states containing Kaons can be very efficiently discriminated, because
the reflected $M^{inv}$ is sufficiently separated from the true signal 
peak. However, this is not true for a pion-muon or pion-electron 
misidentification. 
The separation is slightly better between \doee\ and \dopp.
\item Vertex separation requirements for secondary vertices: \\
Background from light quark production, 
and  of muons from K- and $\pi$- decays within the detector are 
further rejected by exploiting the information of secondary vertices (e.g.
decay length separation, pointing back to primary vertex etc.).
\item Lepton identification (example H1) :\\
Electron identification is possible by using $dE/dx$ measurements
in the drift chambers\km the shower shape analysis in the calorimeter
(and possibly the transition radiators information).
Muons are identified with the instrumented
iron equipped with limited streamer tubes, with the
forward muon system, and in combination with
the calorimeter information.
The momentum has to be above  $\sim 1.5$ to 
$2 \ $ GeV/c  to allow the $\mu$ to reach the instrumented iron.
Thus, the decay  $D^0 \to \mu^+ \mu^-$ suffers from background contributions
by the SCSD mode $D^0 \to \pi^+ \pi^-$, albeit with a known
$BR = 1.6 \cdot 10^{-3} $; here
$\mu$-identification helps extremely well.
An example of background suppression using the particle ID
has been shown in ref.\cite{hera-91a}, 
where a suppression factor of order $O(100)$ has been achieved.
\item Particle ordering methods exploit the fact that
  the decay products of the charmed mesons tend to
  be  the {\it leading} particles in the event (see e.g. \cite{dstar-dis}).
  In the case of observed jets, the charmed mesons are
  expected to carry a large fraction of the jet energy.
\item Event variables such as e.g. the total transverse energy 
   $E_{transverse}$ tend to reflect the difference in  event topology
   between heavy and light quark production processes, and hence
   lend themselves for suppression of light quark background.
\end{itemize}

\subsection{Additional experimental considerations}
\begin{itemize}
\item Further possibilities to enhance overall statistics are the
 usage of inclusive decays (no tagging), where the gain 
 in statistics is expected to be about
  $\frac{ N(all D^0)}{ N(D^0 from D^*)} = 0.61 / 0.21 \approx   3$,
 however on the the cost of higher background contributions.
\item  In the decays  $D^0 \to e e$    or   $D^0 \to \mu e$  one expects
factors of  2 to 5 times better background rejection efficiency.
\item Trigger :
A point to mention separately is the trigger. To be able to 
measure a  BR at the level of $10^{-5}$, the event filtering
process has to start at earliest possible stage.
This should happen preferably at the
first level of the hardware trigger, because it will
not be feasible to store some  $10^{+7}$ events on permanent
storage to dig out the few rare decay candidates.
This point, however, has up to now not yet been thoroughly
studied, let alone been implemented at the 
hardware trigger level.
\end{itemize}

\begin{table}[tb]
\begin{center}
\begin{tabular}{|l|c|c|c|}
\hline
 Mode & BR (90\% C.L.) & Interest & Reference \\
\hline
\hline
 $D^+ \to \pi^+ e^+ e^-$     & $6.6*10^{-5}$ & FCNC  & E791 96  \\
 $D^+ \to \pi^+ \mu^+ \mu^-$ & $1.8*10^{-5}$ & FCNC  & E791 96  \\
 $D^+ \to \pi^+ \mu^+ e^-$ & $3.3*10^{-3}$ & LF& MarkII 90 \\
 $D^+ \to \pi^+ \mu^- e^+$ & $3.3*10^{-3}$ & LF& MarkII 90 \\
\hline
 $D^+ \to \pi^- e^+ e^+$     & $4.8*10^{-3}$ & L& MarkII 90 \\
 $D^+ \to \pi^- \mu^+ \mu^+$ & $2.2*10^{-4}$ & L& E653 95 \\
 $D^+ \to \pi^- \mu^+ e^+$   & $3.7*10^{-3}$ & L+LF& MarkII 90 \\
 $D^+ \to K l l  $ & similar & L+LF& MarkII 90 \\
\hline
 $c \to X \mu^+ \mu^-$ & $1.8*10^{-2}$ & FCNC  & CLEO 88  \\
 $c \to X e^+ e^-$ & $2.2*10^{-3}$ & FCNC  & CLEO 88  \\
 $c \to X \mu^+ e^-$ & $3.7*10^{-3}$ & FCNC  & CLEO 88  \\
\hline
 $D^+ \to \phi K^+ $ & $1.3*10^{-4}$ & DC & E687 95   \\
 $D^+ \to K^+ \pi^+ \pi^- $ & $=6.5*10^{-4}$ & DC & E687 95   \\
 $D^+ \to K^+ K^+ K^- $  & $1.5*10^{-4}$ & DC & E687 95   \\
\hline
 $D^+ \to \mu^+ \nu_{\mu}$ & $7.2*10^{-4}$ & $f_D$ & MarkIII 88 \\
\hline
\hline
 $D_S\to \pi^- \mu^+ \mu^+$ & $4.3*10^{-4}$ & L& E653 95 \\
 $D_S\to K^- \mu^+ \mu^+$ & $5.9*10^{-4}$ & L& E653 95 \\
 $D_S \to \mu^+ \nu_{\mu}$ & $=9 *10^{-4}$ & $f_{D_S}=430$ & BES 95 \\
\hline
\end{tabular}
\caption[Experimental limits on rare $D^+$- and $D_s$-meson decays.]
{Selection of experimental limits at 90\% c.l. 
on rare $D^+$- and $D_s$-meson decays\cite{rf-partbook}
(except where indicated by =).}

\label{tab-ds}
\end{center}
\end{table}



\section{Status of sensitivity in rare charm decays}
 
Some of the current experimental upper limits 
 at 90\% c.l. on the branching ratios of 
rare $D$ decays are summarised in
tables ~\ref{tab-d} and \ref{tab-ds}
according to \cite{rf-partbook}.


Taking the two-body decay $D^0 \to  \mu^+ \mu^-$   to be the
sample case, a comparison of the achievable sensitivity on
the upper limit on branching fraction 
  $B_{D^0 \to  \mu^+ \mu^-}$ at 90\% c.l. is summarized 
in table \ref{tab-comp} for different experiments, 
assuming that NO signal 
events are being detected (see  \cite{rf-grab1}
and  \cite{rf-partbook}). 
Note that the sensitivity reachable at HERA is compatible with
the other facilities, provided the above assumed luminosity is
actually delivered. This does not hold for a
proposed $\tau$-charm factory, which - if ever built and performing
as designed - would exceed all other facilities by at least
two orders of magnitude (\cite{rf-rafetc}).
%


\noindent
The status of competing experiments at other facilities 
is the following :

\noindent
\begin{itemize}
\item SLAC : $e^+e^-$ experiments : Mark-III, MARK-II, DELCO : stopped.
\item CERN : fixed target experiments : ACCMOR, E615, BCDMS, CCFRC : stopped. \\
     LEP-experiments : previously ran at the $Z^0$-peak; 
     now  they continue   
     with increased $\sqrt{s}$, but at a {\it reduced} $\sigma$ for such
     processes;
\item Fermilab (FNAL) : the photoproduction experiments E691/TPS and
      hadroproduction experiments E791 and E653 are
      stopped, with some analyses being finished based on about
      $O(10^5)$ reconstructed events. In the near
      future highly competitive results are to be expected from 
      the $\gamma p$ experiments E687 and 
      its successor E831 (FOCUS), based on an statistics
      of about  $O(10^5)$ and an estimated $10^6$ reconstructed
      charm events, respectively. But also the hadroproduction
      experiment E781 (SELEX) is anticipated to reconstruct some $10^6$ 
      charm events within a few years.
\item DESY : ARGUS $e^+e^-$ : stopped, final papers emerging now. \\
      HERA-B : With a very high cross section of
      $\sigma(p N \to c {\bar c}) \approx 30 \mu$b at
      $\sqrt{s} = 39 $ GeV and an extremely high luminosity,
      a total of up to $10^{12} \ \ c {\bar c}$-events may be
      produced. Although no detailed studies exist so far,
      a sensitivity of order $10^{-5}$ to $10^{-7}$ might be expected,
      depending on the background rates.
\item CESR : CLEO is continuing steadily to collect data, and above all
      is the present leader in 
      sensitivity for many processes (see table \ref{tab-d}).
\item BEPC : BES has collected data at $\sqrt{s}=4.03$ GeV (and 4.14 GeV), 
    and is continuing to do so; BES will become competitive as soon as 
    enough statistics is available, because 
    the background conditions are very favourable.
\item $\tau$-charm factory : The prospects for a facility 
      being built in China (Beijing) are uncertain. 
      If realized, this is going to be the
      most sensitive place to search for rare charm decays. 
      Both, kinematical constraints (e.g. running at the $\psi''(3700)$)
      and the missing background from non-charm induced processes
      will enhance its capabilities.
\end{itemize}

\begin{table}
\begin{center}
\begin{tabular}{|l|c|c|c|c|c|c|c|}
\hline
 $\ $ & SPEAR & BEPC & E691 & LEP & $\tau-c F$ & CLEO & HERA \\
\hline
\hline
$\sigma_{cc}$ (nb) & 5.8 & 8.6 & 500 & 4.5 & 5.8 & 1.3 & $\sigma_{ep}=940$ \\
\hline
$L (pb^{-1}$)    & 9.6 & 30 & 0.5 & 150 & $10^4$ & 3850 & 250\\
\hline
$N_D$                 & $6*10^4$ & $3*10^5$ & $2.5*10^5$ & $10^6$
        & $6*10^7$ & $5 * 10^6$ & $2.4*10^8$\\
\hline
$\epsilon \cdot A$    & 0.4 & 0.5 & 0.25 & 0.05  & 0.5 & 0.1 & 0.06 \\
\hline
$N_{BGND}$  & O(0)  & O(1)  & O(10) & O(10) & O(1) & O(1) & S/N$\approx$1 \\
\hline
$\sigma_{charm} \over \sigma_{total} $ 
                & 1 & 0.4 & 0.01 & 0.1 & 1 & 0.1 & 0.1 \\
\hline
\hline
$B_{D^0 \to  \mu^+ \mu^-}$  & 
  $1.2*10^{-4}$ & $2*10^{-5}$ & $4*10^{-5}$ &
 $5*10^{-5}$ & $5*10^{-8}$ & $3.4 *10^{-5}$ & $2.5*10^{-6}$  \\
\hline
\end{tabular}
\caption[Comparison of sensitivity]
{Comparison of estimated sensitivity to the sample decay mode
$D^0 \to  \mu^+ \mu^-$ for different facilities or experiments.}
\label{tab-comp}
\end{center}
\end{table}


\section{Summary}

$D$-meson decays offer a rich spectrum of interesting physics;  their rare
decays may provide
information on new physics, which is complementary to the 
knowledge stemming from $K$-meson and $B$-decays.
With the prospect of order a few times  $10^8$
produced charmed mesons per year, 
HERA has the potential to contribute substantially to this field.
Further competitive results can be anticipated from the fixed target
experiments at Fermilab or from a possible $\tau$-charm factory.

For the rare decay \domm investigated here we
expect at least an order of magnitude improvement in sensitivity 
over current results (see table given above) for a total integrated luminosity of
\intl = 250 \picb, the limitation here being statistical.
An extrapolation to even higher luminosity is rather difficult
without a very detailed numerical simulation, because
at some (yet unknown) level the background processes will 
become the main limiting factor for the sensitivity, rendering
sheer statistics useless. 
For this, a good tracking resolution, excellent particle
identification (e, $\mu,\ \pi,\ K,\ {\rm p}$) and a high resolution for
secondary vertices is required
to keep the systematics under control, and either to 
unambigously identify a signal of new physics, or to
reach the ultimate limit 
in sensitivity.



\end{document}